%% file: ternary_gapISITarxiv.tex
\documentclass[conference]{IEEEtran}
\usepackage{graphics}
\usepackage{epsfig}

\usepackage[tight]{subfigure}
\usepackage{graphics}

\usepackage{amsthm}
\usepackage{amsmath}
\usepackage{amsfonts}
\usepackage{mathtools}
\mathtoolsset{showonlyrefs=true}
\usepackage{dsfont}
\usepackage{amssymb}
\usepackage{color}
\usepackage{stfloats}
\usepackage{stmaryrd}
\usepackage{cite}
\input{shortcuts.tex}

\begin{document}

\title{On the Finite Length Scaling of Ternary Polar Codes}

\author{\IEEEauthorblockN{Dina Goldin}
\IEEEauthorblockA{School of Electrical and Engineering\\
Tel-Aviv University\\
Tel-Aviv, 6997801 Israel\\
Email: dinagold@post.tau.ac.il}
\and
\IEEEauthorblockN{David~Burshtein}
\IEEEauthorblockA{School of Electrical and Engineering\\
Tel-Aviv University\\
Tel-Aviv, 6997801 Israel\\
Email: burstyn@eng.tau.ac.il}}

\maketitle \setcounter{page}{1}
\begin{abstract}
The polarization process of polar codes over a ternary alphabet is studied. Recently it has been shown that the scaling of the blocklength of polar codes with prime alphabet size scales polynomially with respect to the inverse of the gap between code rate and channel capacity.  However, except for the binary case, the degree of the polynomial in the bound is extremely large. In this work, it is shown that a much lower degree polynomial can be computed numerically for the ternary case. Similar results are conjectured for the general case of prime alphabet size.
\end{abstract}
\begin{IEEEkeywords}
polar codes, scaling, non-binary channels
\end{IEEEkeywords}
\section{Introduction}
Polar codes for transmission over binary discrete memoryless channels (DMCs) were introduced by Arikan \cite{arikan2009channel}, and were further analyzed in \cite{arikan2009rate}. These results were extended to q-ary polarization for an arbitrary prime $q$ in \cite{sasoglu2009polarization,sasoglu2010entropy,karzand2010qsrc}.

For the binary case it was shown that the blocklength required to transmit reliably scales polynomially with respect to the inverse of the gap between code rate and channel capacity \cite{guruswami2013speed,hassani2013finite,goldin2013improved}. This result was recently extended to $q$-ary channels for an arbitrary prime $q$ \cite{guruswami2014entropy} but in the new bound, the degree of this polynomial is extremely large.

In this paper we obtain numerically a much better bound for $q=3$. For that purpose we obtain numerically a lower bound on the size of a basic polarization step which is higher than the one for the binary case. We conjecture similar results for any prime value of the alphabet size, $q$.

\section{Preliminaries}
\subsection{General definitions and results}
We follow the notations of \cite[Lemma 5]{karzand2010qsrc}. For the $q$-ary channel $W\left(y\given x\right)$, we define $W(y)\triangleq (1/q)\sum_{x=0}^{q-1}W\left(y\given x\right)$ and the vector
$
\bv(y)\triangleq\left[v_0(y),v_1(y),\dots,v_{q-1}(y)\right]^T
$
where
\begin{equation}
\forall x\in\left\{0,1,\dots,q-1\right\}: v_x(y)\triangleq\frac{W\left(y\given x\right)}{qW(y)}\:. \label{eq:vxdef}
\end{equation}
Note that $\sum_{x=0}^{q-1}v_x(y)=1$ and the symmetric capacity is
\begin{equation}
I(W)=\sum_y W(y)\left\{1-H\left[\bv(y)\right]\right\} \label{eq:I=sum_y}
\end{equation}
where
\begin{equation}
H\left[\bv(y)\right]\triangleq-\sum_{x=0}^{q-1}v_x(y)\log_q v_x(y)\:. \label{eq:g_def}
\end{equation}
We can rewrite \eqref{eq:I=sum_y} as $I(W)=\sum_G \hW\left(G\right) G$, where
\begin{equation}
\hW\left(G\right)\triangleq\sum_{y:H\left[\bv(y)\right]=1-G} W(y) \label{eq:hat_W_def}
\end{equation}
A basic polarization transformation of a channel $W$ forms two channels, $W^-=W\boxast W$ and $W^+=W\circledast W$. Recall that given two channels, $W_a$ and $W_b$, $W_{a\boxast b} \defined W_a\boxast W_b$ is defined by
\begin{equation}
W_{a\boxast b}\left(y_1,y_2\given u\right) \defined \frac{1}{q}\sum_{u'=0}^{q-1}W_b\left(y_2\given u'\right)W_a\left(y_1\given u+u'\right)
\end{equation}
Hence $W_{a\boxast b}\left(y_1,y_2\right)=W_a\left(y_1\right)W_b\left(y_2\right)$ and \cite[Proof of Lemma 6]{karzand2010qsrc}
\begin{equation}
v_{a\boxast b,u}\left(y_1,y_2\right)=\sum_{u'=0}^{q-1}v_{b,u'}\left(y_2\right)v_{a,u+u'}\left(y_1\right) \label{eq:v^-karzand}
\end{equation}
which can be rewritten as
\begin{equation}
	\bv_{a\boxast b}\left(y_1,y_2\right)=\bv_b\left(y_2\right)\star \bv_a\left(y_1\right) \label{eq:v_star}
\end{equation}
where $\star$ denotes circular cross-correlation with period $q$.
Defining
\begin{equation} g\left(G_1,G_2\right)\triangleq1-\min_{\substack{H\left[\bv_a\left(y_1\right)\right]=1-G_1\\ H\left[\bv_b\left(y_2\right)\right]=1-G_2}} H\left[\bv_b\left(y_2\right)\star \bv_a\left(y_1\right)\right]\label{eq:hat_g_def}
\end{equation}
we obtain
\begin{align}
&I\left(W_{a\boxast b}\right)=\sum_{y_1,y_2}W_{a\boxast b}\left(y_1,y_2\right)\left\{1-H\left[\bv_{a\boxast b}\left(y_1,y_2\right)\right]\right\}\\
&\quad\le \sum_{G_1,G_2}\sum_{\substack{y_1:H\left[\bv_a\left(y_1\right)\right]=1-G_1\\ y_2:H\left[\bv_b\left(y_2\right)\right]=1-G_2}} W_a\left(y_1\right)W_b\left(y_2\right) g\left(G_1,G_2\right)\\
&\quad=\sum_{G_1,G_2}\hW_a\left(G_1\right)\hW_b\left(G_2\right) g\left(G_1,G_2\right)
\end{align}
where the first equality is an application of \eqref{eq:I=sum_y}, the inequality follows from \eqref{eq:v_star}, \eqref{eq:hat_g_def} and $W_{a\boxast b}\left(y_1,y_2\right)=W_a\left(y_1\right)W_b\left(y_2\right)$, and \eqref{eq:hat_W_def} yields the last equality.
If $ g\left(G_1,G_2\right)$ is concave in $G_1$ and separately, not necessarily jointly, in $G_2$
\begin{multline}
I\left(W_{a\boxast b}\right)\le g\left[\sum_{G_1}\hW_a\left(G_1\right)G_1,\sum_{G_2}\hW_a\left(G_2\right)G_2\right]\\= g\left[I\left(W_a\right),I\left(W_b\right)\right] \label{eq:Iconv}
\end{multline}
and since $W^-=W\boxast W$, $I\left(W^-\right)\le g\left[I(W),I(W)\right]$.
If $ g\left(G_1,G_2\right)$ is not concave in $G_1$ and in $G_2$, we can replace it with a concave upper-bound, and \eqref{eq:Iconv} will remain true.

Note that by \eqref{eq:v_star}, $v_{a\boxast b,u}\left(y_1,y_2\right)=v_{b\boxast a,-u}\left(y_2,y_1\right)$, where the subtraction is modulo $q$. Combining this with \eqref{eq:hat_g_def} yields $ g\left(G_1,G_2\right)= g\left(G_2,G_1\right)$.
\subsection{Proved results about the QSC channel}
A $q$-ary symmetric channel (QSC) $W\left(y\given x\right)$ with error probability $p$ is defined by \begin{equation}
W\left(y\given x\right)=\left\{
\begin{array}{ll}
1-p & y=x\\
p/(q-1) & y\ne x\;.
\end{array}
\right.\end{equation}
Although the QSC channel does not maximize \eqref{eq:hat_g_def} for some pair $(G_1,G_2)$, we observed that for $q=3$ it provides an excellent approximation to the maximum, and we conjecture that this holds true for any prime $q$.
\begin{lemma} \label{lemma:QSC}
If $W_a$ and $W_b$ are QSC channels, then $W_{a\boxast b}$ is a QSC channel as well.
Furthermore, $I\left(W_{a\boxast b}\right)= g_{QSC}\left[I\left(W_a\right),I\left(W_b\right)\right]$ for
\begin{multline}
 g_{QSC}\left(G_1,G_2\right)\triangleq 1-h_q\bigg[h_q^{-1}\left(1-G_1\right)+h_q^{-1}\left(1-G_2\right)\\ \left.-\frac{q}{q-1}h_q^{-1} \left(1-G_1\right)h_q^{-1}\left(1-G_2\right)\right] \label{eq:hat_g_QSC}
\end{multline}
with $h_q(p)\defined -(1-p)\log_q(1-p)-p\log_q\left(\frac{p}{q-1}\right)$ and $h_q^{-1}$ is the inverse of $h_q$, that yields values in $\left[0,\frac{q-1}{q}\right]$.
\end{lemma}
The proof of this Lemma is a straightforward application of \eqref{eq:vxdef} and \eqref{eq:v_star}.
\begin{lemma} \label{lemma:QSCLag3}
Using QSC channels $W_a$ and $W_b$ yields an extreme point in the Lagrangian related to \eqref{eq:hat_g_def} for $G_1,G_2>0$.
\end{lemma}
The proof of this Lemma is also straightforward.
\section{Analysis and Numerical Results}
Observe the similar to \eqref{eq:hat_g_def} problem
\begin{equation} \tilde{g}\left(G_1,G_2\right)\triangleq1-\min_{\substack{H\left[\bv_a\left(y_1\right)\right]\ge 1-G_1\\H\left[\bv_b\left(y_2\right)\right]\ge 1-G_2}} H\left[\bv_b\left(y_2\right)\star \bv_a\left(y_1\right)\right]\label{eq:tilde_g_def}
\end{equation}
First, we prove the following.
\begin{lemma} \label{lemma:joint}
Define $f\left(\bu\right)\triangleq \min_{H\left(\bv\right)\ge 1-G} H\left(\bu\star\bv\right)$. Then, $f\left(\bu\right)$ is concave.
\end{lemma}
\begin{IEEEproof}
By definition, $f\left(\bu_0\right)\triangleq \min_{H\left(\bv\right)\ge 1-G} H\left(\bu_0\star\bv\right)$ and $f\left(\bu_1\right)\triangleq \min_{H\left(\bv\right)\ge 1-G} H\left(\bu_1\star\bv\right)$. Then
\begin{align}
&f\left(\alpha\bu_0+(1-\alpha)\bu_1\right)\\
&\quad=\min_{H\left(\bv\right)\ge 1-G} H\left(\alpha\bu_0\star\bv+(1-\alpha)\bu_1\star\bv\right)\\
&\quad\ge \min_{H\left(\bv\right)\ge 1-G}\left[\alpha H\left(\bu_0\star\bv\right)+(1-\alpha)H\left(\bu_1\star\bv\right)\right]\\
&\quad\ge\alpha \min_{H\left(\bv\right)\ge 1-G} H\left(\bu_0\star\bv\right)+ (1-\alpha)\min_{H\left(\bv\right)\ge 1-G} H\left(\bu_1\star\bv\right)\\
&\quad=\alpha f\left(\bu_0\right)+(1-\alpha)f\left(\bu_1\right)
\end{align}
where the first inequality follows from concavity of $H$, and the added degree of freedom to the minimization yields the second inequality.
\end{IEEEproof}
Since the constraints in this problem form a convex region, and by Lemma \ref{lemma:joint} we minimize a concave function, $f(\bu)$, the result is obtained on the boundary of the convex region, and $\tilde{g}= g$. Note that Lemma~\ref{lemma:joint} enables us to compute $g$ efficiently using known algorithms for concave minimization over a convex region \cite{hoffman1981method}. This algorithm generates linear programs whose solutions minimize the convex envelope of the original function over successively tighter polytopes enclosing the feasible region. As the polytopes become more complex and more tight, the generated solution becomes more precise.

We can now prove the following.
\begin{lemma} \label{lemma:prop}
$g\left(G_1,G_2\right)$ has the following properties:
\begin{enumerate}
\item $ g\left(x_1,y_1\right)\le  g\left(x_2,y_2\right)$ for $x_1\le x_2$ and $y_1\le y_2$. \label{state:1}
\item $ g\left(1,G_2\right)=G_2$ \label{state:ends}
\item $ g\left(G_1,G_2\right)\le \min\left(G_1,G_2\right)$. \label{state:2}
\item $\lim_{x\rightarrow 1}\frac{\partial  g\left(x,G_2\right)}{\partial x}=0$ \label{state:3}
\end{enumerate}
\end{lemma}
\begin{IEEEproof}
Since $x_1\le x_2$ and $y_1\le y_2$, the constraints for $\tilde{g}\left(x_1,y_1\right)$ are tighter than the constraints for $\tilde{g}\left(x_2,y_2\right)$. Since it is a maximization problem ($1-\min$), the maximum for $\left(x_1,y_1\right)$ would be smaller than the maximum for $\left(x_2,y_2\right)$, i.e. $\tilde{g}\left(x_1,y_1\right)\le \tilde{g}\left(x_2,y_2\right)$. Since $\tilde{g}= g$, statement \ref{state:1} follows. Statement \ref{state:ends} follows since for $G_1=1$, $\bv_a\left(y_1\right)$ is a circular permutation of $\left[1,0,\dots,0\right]^T$, so by \eqref{eq:g_def} and \eqref{eq:v_star}, $H\left[\bv_{a\boxast b}\left(y_1,y_2\right)\right]=H\left[\bv_b\left(y_2\right)\right]$ Now, $ g\left(G_1,G_2\right)\le g\left(G_1,1\right)=G_1$ and $ g\left(G_1,G_2\right)\le  g\left(1,G_2\right)=G_2$, which yields statement \ref{state:2}.
Since \eqref{eq:hat_g_def} is a maximization problem, Lemma \ref{lemma:QSCLag3} yields that $ g\left(x,G_2\right)\ge g_{QSC}\left(x,G_2\right)$, where $g_{QSC}$ is defined in \eqref{eq:hat_g_QSC}.
By parts 1) and 2), $g(x,G_2) \le g(1,G_2) = G_2 = g_{QSC}(1,G_2)$.
Also, straightforward calculations show that $\lim_{x\rightarrow 1}\frac{\partial g_{QSC}\left(x,G_2\right)}{\partial x}=0$.
Combining the above yields statement \ref{state:3}.
\end{IEEEproof}
Next, we calculate $g\left(G_1,G_2\right)$ for $G_1,G_2\approx 0$ and for $G_1,G_2\approx 1$. To simplify the notation, we will denote $\bv_a\left(y_1\right)=\bv_a = \left[v_{a,0},v_{a,1},\dots, v_{a,q-1}\right]^T$, $\bv_b\left(y_2\right)=\bv_b=\left[v_{b,0},v_{b,1},\dots, v_{b,q-1}\right]^T$ and $\bv_{a\boxast b}\left(y_1,y_2\right)=\bv_t=\left[v_{t,0},v_{t,1},\dots,v_{t,q-1}\right]^T$.
\begin{lemma} \label{lemma:smallG1G2}
For sufficiently small values of $G_1$ and $G_2$ and $q=3$, $g\left(G_1,G_2\right)=\ln 3\cdot G_1G_2$.
\end{lemma}
\begin{IEEEproof}
Consider \eqref{eq:hat_g_def}.  For $G_2$ sufficiently small, $v_{b,i} = 1/q + \epsilon_i$ where $\epsilon_i$ are sufficiently small and $\sum_{i=0}^{q-1}\epsilon_i=0$. Using Taylor's approximation, and $\gamma\triangleq q/(2\ln q)$, $H\left[\bv_b\right]=1-\gamma\sum_{i=0}^{q-1}\epsilon_i^2$.
We shall first solve the minimization problem in \eqref{eq:hat_g_def} for a fixed $\bv_a$ and $G_2\approx 0$, so
$v_{t,i}=1/q+\sum_{k=0}^{q-1}\epsilon_k v_{a,i+k}$ and
$
H\left[\bv_t\right]=1-\gamma\sum_{i=0}^{q-1}\left(\sum_{k=0}^{q-1}\epsilon_k v_{a,i+k}\right)^2
$.
Hence,
$g\left(G_1,G_2\right)=\gamma\max\sum_{i=0}^{q-1}\left(\sum_{k=0}^{q-1}\epsilon_k v_{a,i+k}\right)^2=\gamma \max\bepsilon^T A\bepsilon$ s.t. $\bepsilon^T\bepsilon=G_2/\gamma$ and $\sum_{i=0}^{q-1}\epsilon_i=0$.
Here $\bepsilon=\left[\epsilon_0,\dots,\epsilon_{q-1}\right]^T$ and $A = \sum_{i=0}^{q-1} {\bf v}_{a,i}{\bf v}_{a,i}^T$ where ${\bf v}_{a,i}$ is a cyclic shift by $i$ of ${\bf v}_a$. Hence,
\begin{equation}
g\left(G_1,G_2\right)=G_2\max\bepsilon^T A\bepsilon \text{ s.t. }\bepsilon^T\bepsilon=1,\sum_{i=0}^{q-1}\epsilon_i=0\;. \label{eq:quad_form}
\end{equation}
 Note that $A$ is a circulant matrix, and for $q=3$
$$
a_{i,j}=\left\{\begin{array}{ll}
\sum_{k=0}^2 v_{a,k}^2 & i=j\\
\sum_{k=0}^2 v_{a,k}v_{a,k+1} & i\ne j
\end{array}\right.
$$
so $A$ has only two eigenvalues: $\lambda_1=1$ and $\lambda_2= \sum_{k=0}^2 v_{a,k}^2- \sum_{l=0}^2v_{a,k}v_{a,k+1}<\lambda_1$. The eigenvector associated with $\lambda_1$ is $\bu_1=c[1,1,1]^T$ so the linear constraint can be expressed as $\bepsilon^T\bu_1=0$. Following \cite[page 411, Th. 7]{lay2012linear}, the solution to \eqref{eq:quad_form} is $G_2\lambda_2$. The eigenvector associated with $\lambda_2$ is $\bepsilon=c\left[1,-0.5,-0.5\right]^T$, making $W_b$ a QSC channel. Substituting it into  \eqref{eq:quad_form} yields
\begin{equation}
g\left(G_1,G_2\right)= G_2 \left(\sum_{i=0}^2 v_{a,i}^2- \sum_{i=0}^2v_{a,i}v_{a,i+1}\right) \label{eq:HsmallG}
\end{equation}
For $G_1\approx 0$, $v_{a,i}=1/3+\delta_i$, $\sum_{i=0}^2\delta_i=0$ and $\sum_{i=0}^2 v_{a,i}^2- \sum_{i=0}^2v_{a,i}v_{a,i+1}=\sum_{i=0}^2\delta_i^2-\sum_{i=0}^2\delta_i\delta_{i+1}$.

Since $\sum_{i=0}^2\delta_i=0$, $\sum_{i=0}^2\delta_i^2=2\left(\delta_1^2+\delta_2^2+\delta_1\delta_2\right)$ and
$\sum_{i=0}^2\delta_i\delta_{i+1}=-\left(\delta_1^2+\delta_2^2+\delta_1\delta_2\right)$. Therefore, $\sum_{i=0}^2\delta_i^2-\sum_{i=0}^2\delta_i\delta_{i+1}=1.5\sum_{i=0}^2\delta_i^2=\frac{3G_1}{2\gamma}$.
Combining this with \eqref{eq:HsmallG} yields the stated result.
\end{IEEEproof}
\begin{lemma} \label{lemma:largeG1G2}
For $G_1$ and $G_2$ sufficiently close to $1$, and $q=3$, $g(G_1,G_2) = G_1 + G_2 - 1$
\end{lemma}
\begin{IEEEproof}
Consider \eqref{eq:hat_g_def}. For $G_1$ sufficiently close to $1$, we can assume without loss of generality that $v_{a,i}= \delta_i$, $i=1,\dots,q-1$, where $\delta_i$ are small, and $v_{a,0} = 1-\sum_{i=1}^{q-1}\delta_i$.
Similarly, for $G_2$ sufficiently close to $1$, we can assume without loss of generality that
$v_{b,i}= \epsilon_i$, $i=1,\dots,q-1$, where $\epsilon_i$ are small, and $v_{b,0} = 1-\sum_{i=1}^{q-1}\epsilon_i$.
Now, $1 - G_1 = H\left[ \bv_a \right] = -\sum_{i=1}^{q-1}\delta_i\log_q\delta_i$ and $1 - G_2 = H\left[ \bv_b \right] = -\sum_{i=1}^{q-1}\epsilon_i\log_q\epsilon_i$.
For $G_1$ and $G_2$ sufficiently close to $1$, $\bv_t \approx [1-\delta_1-\delta_2-\epsilon_1-\epsilon_2,\delta_1+\epsilon_2+\epsilon_1\delta_2,\delta_2+\epsilon_1+\epsilon_2\delta_1]^T$. Hence, $H\left[ \bv_t \right] = -(\delta_1+\epsilon_2+\epsilon_1\delta_2)\log_3(\delta_1+\epsilon_2+\epsilon_1\delta_2)-(\delta_2+\epsilon_1+\epsilon_2\delta_1)\log_3(\delta_2+\epsilon_1+\epsilon_2\delta_1)$.
Now, our main observation is that for $a,b,c$ small, $-(a+b+c)\log(a+b+c) \approx -a\log a -b\log b -c\log c$. Applying this observation and $-\epsilon\delta\log(\epsilon\delta)<<-\epsilon\log\epsilon-\delta\log\delta$ for small $\epsilon$ and $\delta$ yields $H\left[ \bv_t \right] \approx 2 - G_1 - G_2$ so that $g(G_1,G_2) \approx G_1 + G_2 -1$. 
\end{IEEEproof}
Note that the same proof applies for a general $q$.

\begin{figure}
\centering
\includegraphics[width=\columnwidth]{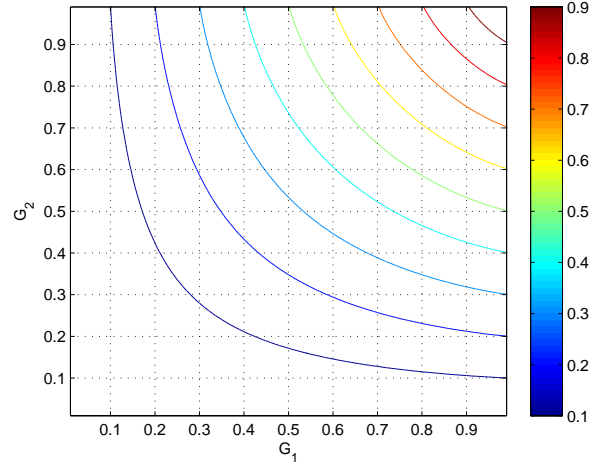}
\caption{Numerically calculated $g\left(G_1,G_2\right)$ for $q=3$}
\label{fig:g_num}
\end{figure}

We calculated the actual value of $g$ numerically. We calculated $ g \left(0.01n,0.01m\right)$ for $q=3$, $n=1,2,\dots,99$ and $m=1,2,\dots,99$.
In Figure~\ref{fig:g_num} we plot the contour of this function. This figure shows that $ g\left(G_1,G_2\right)= g\left(G_2,G_1\right)$ as noted above, and, as proved in Lemma \ref{lemma:prop}, $g\left(1,G_2\right)=G_2$.

\begin{figure}
\centering
\includegraphics[width=\columnwidth]{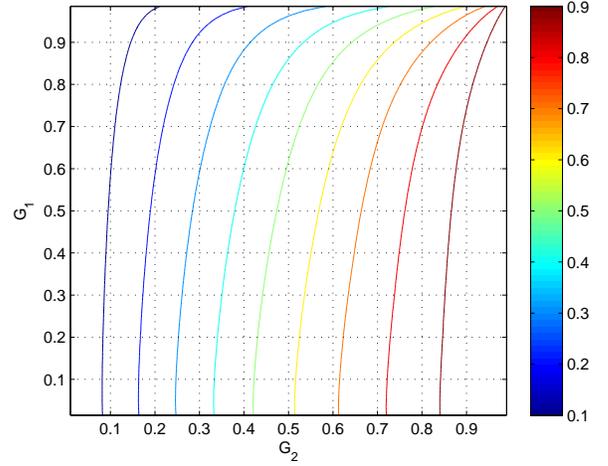}
\caption{Numerically calculated $\frac{\partial g\left(G_1,G_2\right)}{\partial G_1}$ for $q=3$}
\label{fig:dg_num}
\end{figure}

\begin{figure}
\centering
\includegraphics[width=\columnwidth]{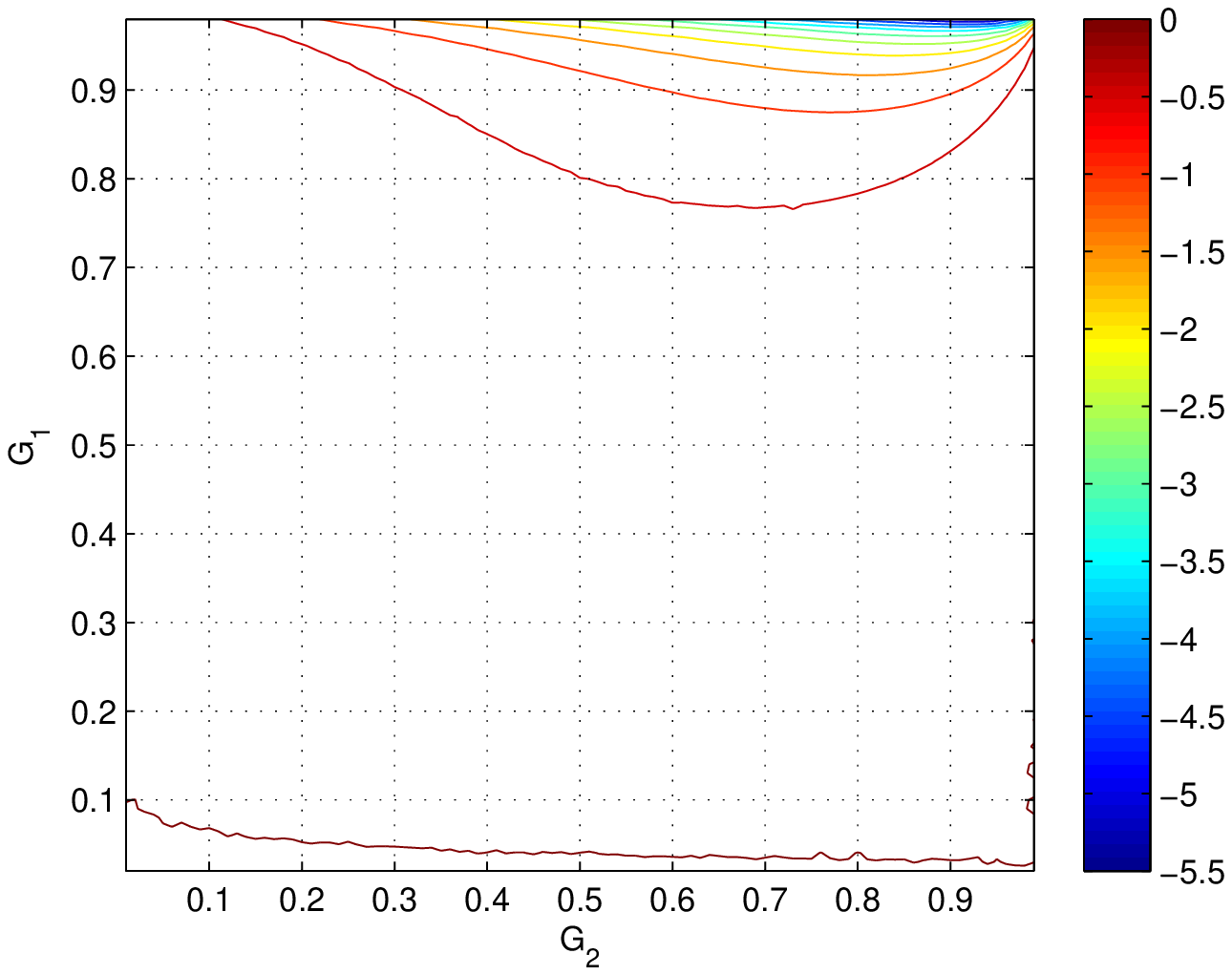}
\caption{Numerically calculated $\frac{\partial^2 g\left(G_1,G_2\right)}{\partial G_1^2}$ for $q=3$}
\label{fig:d2g_num}
\end{figure}

Plotting the numeric $\frac{\partial g\left(G_1,G_2\right)}{\partial G_1}$ in Figure \ref{fig:dg_num} shows that $ g\left(G_1,G_2\right)$ is increasing in $G_1$ (and by symmetry, in $G_2$), as proved in Lemma \ref{lemma:prop}.
Next, using the calculated points, we estimate $\frac{\partial^2 g\left(x,G_2\right)}{\partial x^2}$. This estimated second derivative is shown in Figure \ref{fig:d2g_num}, suggesting the following conjecture (since the bottom line represents $\frac{\partial^2 g\left(G_1,G_2\right)}{\partial G_1^2}=0$, so below it $\frac{\partial^2 g\left(G_1,G_2\right)}{\partial G_1^2}>0$ and $\frac{\partial^2 g\left(G_1,G_2\right)}{\partial G_1^2}<0$ above that line):
\begin{property} \label{prop:g(G1,G2)}
$ g\left(G_1,G_2\right)$ is concave in $G_1$ (and by symmetry, in $G_2$), except for small values of $G_1$ and $G_2$. In other words, 
for each $G_2\in(0,1)$ there exists $x^*$ s.t. $\frac{\partial^2 g\left(x,G_2\right)}{\partial x^2}$ is positive for $x<x^*$ and negative for $x>x^*$.
\end{property}
Therefore, the convex hull of $ g\left(G_1,G_2\right)$ for a given $G_2$ is
\begin{equation}
\max_{x\in\left[G_1,1\right]}\frac{G_1}{x} g\left(x,G_2\right)=
\left\{\begin{array}{ll}
\frac{G_1}{G_1^*} g\left(G_1^*,G_2\right) & G_1\le G_1^*\\
 g\left(G_1,G_2\right) & G_1\ge G_1^*
\end{array}\right.
\end{equation}
where $G_1^*=\argmax_{x\in[0,1]}\frac{ g\left(x,G_2\right)}{x}$. Finding $G_1^*$ is equivalent to solving $\frac{\partial g\left(x,G_2\right)}{\partial x}=\frac{ g\left(x,G_2\right)}{x}$ s.t. $\frac{\partial^2 g\left(x,G_2\right)}{\partial x^2}<0$, i.e. finding a tangent to $ g\left(x,G_2\right)$ at $x$ s.t. $\frac{\partial^2 g \left(x,G_2\right)}{\partial x^2}<0$, that passes through $\left(0,0\right)$.
\begin{lemma} \label{lemma:one-sol}
If Property \ref{prop:g(G1,G2)} holds, the problem $x\cdot\frac{\partial g \left(x,G_2\right)}{\partial x}= g \left(x,G_2\right)$ s.t. $\frac{\partial^2 g \left(x,G_2\right)}{\partial x^2}<0$ has a single solution.
\end{lemma}
The proof of this Lemma follows from analysis of $x\cdot\frac{\partial g \left(x,G_2\right)}{\partial x}- g \left(x,G_2\right)$.

\begin{figure}
\centering
\includegraphics[width=\columnwidth]{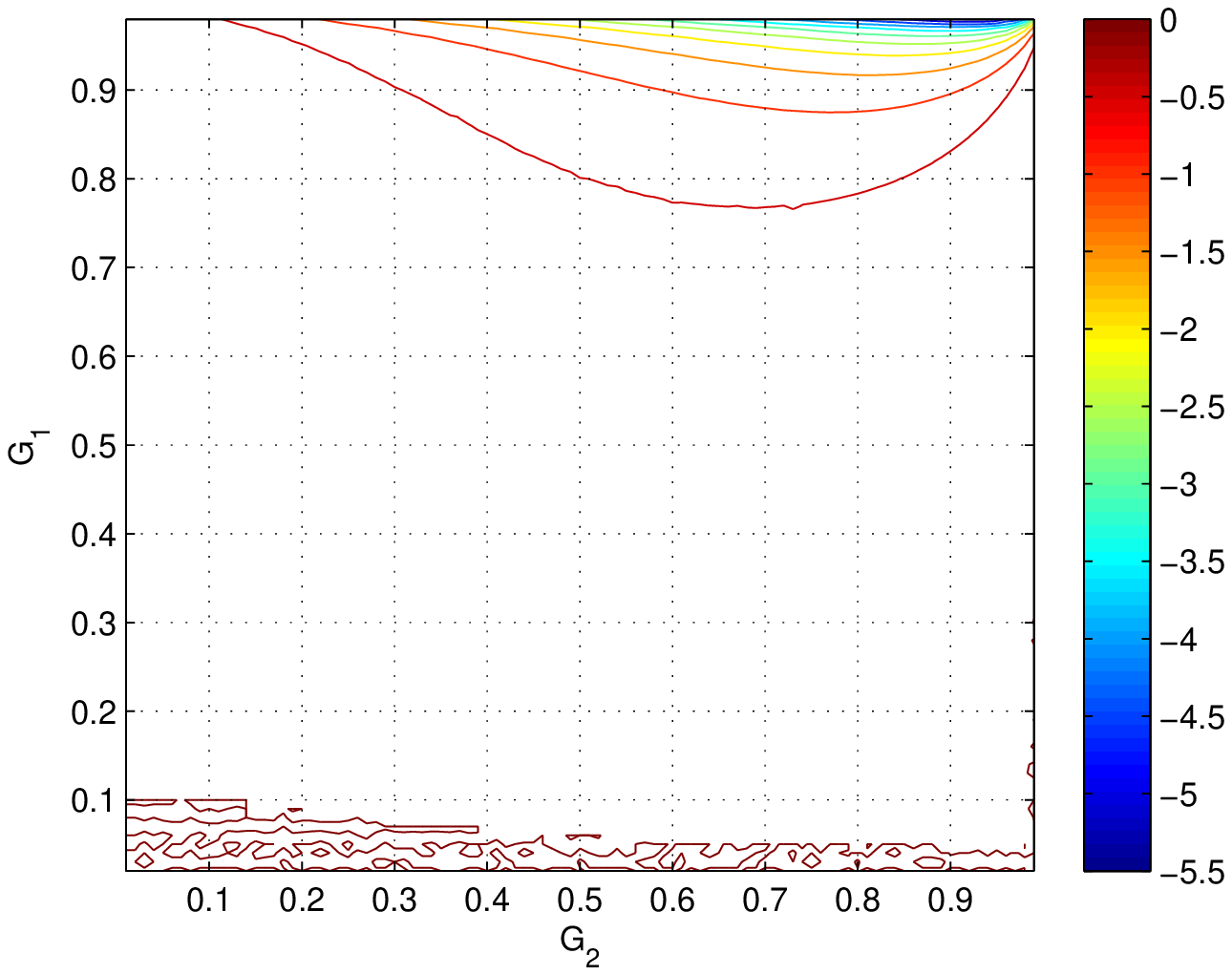}
\caption{Numerically calculated $\frac{\partial^2g^*\left(G_1,G_2\right)}{\partial G_1^2}$ for $q=3$}
\label{fig:d2g_conv_num}
\end{figure}

\begin{figure}
\centering
\includegraphics[width=\columnwidth]{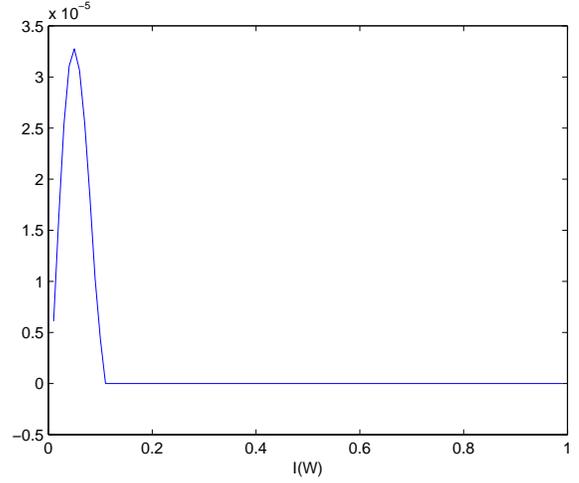}
\caption{Numerically calculated $\epsilon_l(I(W))-\epsilon_l^*(I(W))$ for $q=3$}
\label{fig:g_conv-g_num}
\end{figure}

\begin{figure}
	\centering
		\includegraphics[width=\columnwidth]{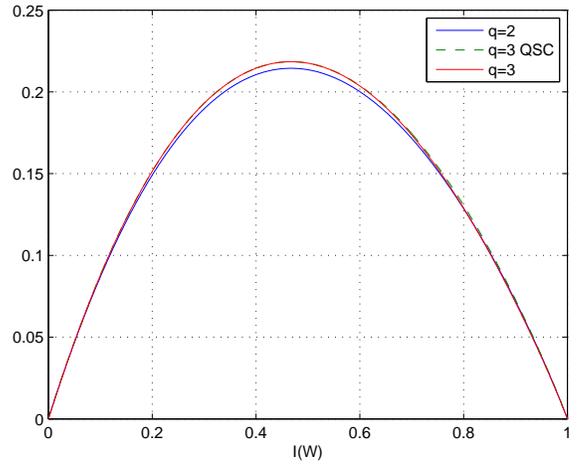}
\caption{The lower bound on $I\left(W^+\right)-I\left(W\right)$, which is also a lower bound on $I\left(W\right)-I\left(W^-\right)$, for different values of $q$, and for the QSC channel}
	\label{fig:min_step31}
\end{figure}

However, we want an upper bound on $ g\left(G_1,G_2\right)$ that would be concave in $G_1$ and $G_2$. Similarly to the case of fixed $G_2$,
\begin{equation} \label{eq:hat_g*}
g^*\left(G_1,G_2\right)=\max_{\substack{x_1\in\left[G_1,1\right]\\x_2\in\left[G_2,1\right]}}\frac{G_1G_2}{x_1x_2} g\left(x_1,x_2\right)
\end{equation}
Clearly, $g^*\left(G_1,G_2\right)\ge g\left(G_1,G_2\right)$ and Figure \ref{fig:d2g_conv_num} shows that $g^*\left(G_1,G_2\right)$ is concave in $G_1$ and in $G_2$ (the lines at the bottom of the figure stand for the area where $\frac{\partial^2 g\left(G_1,G_2\right)}{\partial G_1^2}=0$.
\begin{proposition} \label{prop:eps_l}
There exists $\epsilon_l^*(x)$ s.t.
$
I\left(W^-\right)+\epsilon_l^*\left[I(W)\right]\le I(W) \le I\left(W^+\right)-\epsilon_l^*\left[I(W)\right]
$.
\end{proposition}
\begin{IEEEproof}
Set $\epsilon_l^*(x)=x-g^*(x,x)$, where $g^*(x,x)$ was defined in \eqref{eq:hat_g*}. Recalling that $I\left(W^-\right)\le g^*\left[I(W),I(W)\right]$ and $I\left(W^-\right)+I\left(W^+\right)=2I(W)$ yields the stated result.
\end{IEEEproof}
The minimal polarization step size is $\epsilon_l^*(x)$ rather than $\epsilon_l(x)=x- g(x,x)$. However, $\epsilon_l(x)-\epsilon_l^*(x)$ is very small, as seen in Figure \ref{fig:g_conv-g_num}, so we can use $\epsilon_l(x)$, which is easier to calculate.
In Figure \ref{fig:min_step31} we plot $\epsilon_l(x)$ for different values of $q$, and see that for $q=3$, $\epsilon_l(x)$ is close, but not equal to
$
\epsilon_{l,QSC}(x)=x+h_q\left\{h_q^{-1}\left(1-x\right)\left[2-\frac{q}{q-1}h_q^{-1}\left(1-x\right)\right]\right\}-1 $
which is marked as ``$q=3$ QSC''. From Lemma \ref{lemma:smallG1G2}, $\epsilon_l(x)\approx x-\ln 3\cdot x^2$ for $x\rightarrow 0$, so $\lim_{x\rightarrow 0}\frac{\partial\epsilon_l(x)}{\partial x}=1$, as seen in Figure \ref{fig:min_step31}. Lemma \ref{lemma:largeG1G2} yields $\epsilon_l(x)\approx 1-x$ for $x\rightarrow 1$, as can be seen in Figure \ref{fig:min_step31}.
Note that for $q=2$, we would obtain the same $\epsilon_l(x)=\epsilon_l^*(x)=\epsilon_{l,QSC}(x)$ as in \cite{hassani2013finite}.

Given some function $f_0(x)$, defined over $[0,1]$ s.t. $f_0(x)>0$ for $x\in(0,1)$, and $f_0(0)=f_0(1)=0$, we define $f_k(x)$ for $k=1,2,\ldots$ recursively as follows,
\begin{equation}
f_k(x)\triangleq\sup_{\epsilon_l(x)\le \epsilon\le\epsilon_h(x)} \frac{f_{k-1}(x+\epsilon)+f_{k-1}(x-\epsilon)}{2}
\end{equation}
where $\epsilon_l(x)=x- g(x,x)$ and $\epsilon_h(x)=\min(x,1-x)$.

\begin{figure}
\centering
\includegraphics[width=\columnwidth]{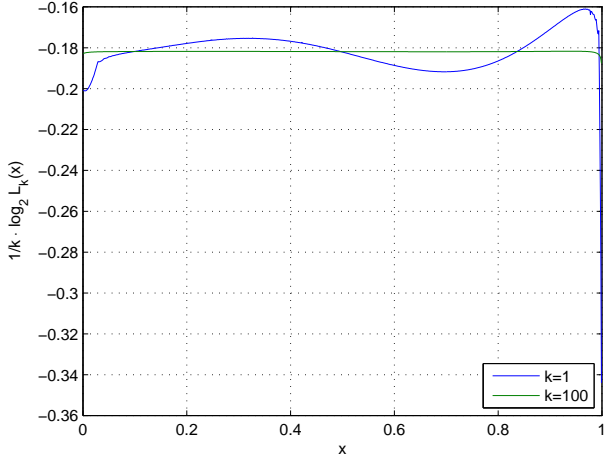}
\caption{A plot of $\frac{1}{k}\log L_k(x)$ for $k=1$ and $k=100$, $q=3$ and $f_0(x)=\left(0.26x^2+1\right)x^{0.8}(1-x)^{0.6}$. The functions $L_k(x)$ were calculated numerically.} \label{fig:Lx3}
\end{figure}

Define
$L_k(x) = \frac{f_k(x)}{f_0(x)}$ and $L_k = \sup_{z\in(0,1)} L_k(z)$.
With the definition of $f_k(x)$, $\sqrt[k]{L_k} \le L_1$ still holds as in \cite{goldin2013improved}.
Similarly to \cite[Equation (11)]{goldin2013improved} we have, for an integer $0<k<n$,
\begin{equation}
\rE\left[f_0\left(I_n\right)\right] \le
\left(\frac{L_1}{\sqrt[k]{L_k}}\right)^{k-1} \cdot
\left(\sqrt[k]{L_k}\right)^n \cdot f_0\left[I(W)\right] \;. \label{eq:HassaniIk}
\end{equation}
Similarly to~\cite{goldin2013improved} we define $J_n \defined \min(I_n,1-I_n)$.
Using
$
f_0(z)=\left(0.26x^2+1\right)x^{0.8}(1-x)^{0.6} 
$
similarly to \cite[Lemma 3]{goldin2013improved}, we obtain
$
P\left(J_n>\delta\right)\le \frac{\alpha_1}{2\delta} \cdot 2^{-0.1817n}
$.
As can be seen in Figure \ref{fig:Lx3}, numerical calculations yield $L_1=2^{-0.161}$ and, $\sqrt[100]{L_{100}} =2^{-0.1817}$.
A plot of $\frac{1}{k}\log_2 L_k$ as a function of $k$ for $q=3$ and $f_0(z)=\left(0.26x^2+1\right)x^{0.8}(1-x)^{0.6}$ 
shows a convex decreasing function, similar to \cite[Fig. 3]{goldin2013improved}, suggesting that it is reasonable to expect that for this particular $f_0(z)$, using $k=100$ is already a good choice for \eqref{eq:HassaniIk} (i.e., we cannot improve much by using an higher value of $k$).
Similarly to \cite[Lemma 4]{goldin2013improved} we have the following. If
$
P\left[\omega\in\Omega \: : \: I_n(\omega) \not\in (\delta,1-\delta) \: \: \forall n\ge m_0 \right]
\ge 1-\epsilon
$
for some integer $m_0$, $0<\epsilon<1$ and $\delta<1/3$. Then
\begin{align}
P\left(\omega\in\Omega \: : \: I_n(\omega)\ge1-\delta \: \: \forall n\ge m_0 \right)
&\ge I(W)-\epsilon\\
P\left(\omega\in\Omega \: : \: I_n(\omega)\le\delta \: \: \forall n\ge m_0 \right)
&\ge 1-I(W)-\epsilon \;.
\end{align}
The proof is essentially the same as the proof of \cite[Lemma 4]{goldin2013improved}, with $I_n$ replacing $1-Z_n$.
Finally, we can obtain a result similar to \cite[Theorem 1]{goldin2013improved}. We use essentially the same proof but with the following modification. First we obtain a result similar to \cite[Equation (25)]{goldin2013improved} using the same approach:
$
P\left(\omega\in\Omega \: : \: I_n(\omega) \ge 1-\delta \: \: \forall n\ge m_0 \right) \ge I(W)-\left(\frac{\alpha_1}{2\delta}\right)\cdot \frac{2^{-\rho m_0}}{1-2^{-\rho}} \;.
$
Then we combine it with~\cite[Equation (2)]{arikan2009channel} to obtain,
$
P\left(\omega\in\Omega \: : \: Z_n(\omega)\le\zeta \: \: \forall n\ge m_0 \right)
\ge
I(W)-\left(\frac{\alpha_1}{\zeta^2}\right)\cdot \frac{2^{-\rho m_0}}{1-2^{-\rho}}
$
and proceed with the derivation in \cite[Theorem 1]{goldin2013improved}. Since $\rho=0.1817$, $1+1/\rho=6.504$, we claim the following result
\begin{proposition}\label{theo:hassani17}
Suppose that we wish to use a polar code with rate $R$ and blocklength $N$ to transmit over a binary-input channel, $W$, with block error probability at most $P_e^0$. Then it is sufficient to set
$
N = \frac{\beta}{\left( I(W)-R \right)^{6.504}}
$
(or larger) where $\beta$ is a constant that depends only on $P_e^0$.
\end{proposition}
\section{Future Research}
In this paper we showed numerically that for the case where $q=3$ we can obtain an improved lower bound on $I(W)-I(W^-)$ compared to the binary ($q=2$ case). Consequently we can predict a much better scaling law of the blocklength with respect to $I(W)-R$ compared to the results in \cite{guruswami2014entropy}. It is interesting to continue this study for other values of prime $q$.
\bibliographystyle{IEEEtran}
\bibliography{bibliography}
\newpage
\section*{Appendix: Supplementary Material}
\subsection{Proof of Lemma \ref{lemma:QSC}} \label{appendix:lemmaQSC}
Assume that $W_a$ and $W_b$ are QSC channels with error probabilities $p_a$ and $p_b$, respectively.
Then, for all $y=0,1,\dots,q-1$, $\bv_a\left(y\right)$ and $\bv_a\left(y\right)$ are circular shifts of $\tilde\bv_a$ and $\tilde\bv_b$, respectively, where
\begin{align}
\tilde\bv_a&\triangleq \left[1-p_a,p_a/(q-1),p_a/(q-1),\dots,p_a/(q-1)\right]^T\\
\tilde\bv_b&\triangleq \left[1-p_b,p_b/(q-1),p_b/(q-1),\dots,p_b/(q-1)\right]^T\;.
\end{align}
Since for the QSC case, all $\bv(y)$ vectors are shifts of some $\tilde\bv$, if $W$ is a QSC channel, $I(W)=1-H\left(\tilde\bv\right)$. This means
\begin{align}
I\left(W_a\right)&=1-h_q\left(p_a\right) \label{eq:BSCa}\\
I\left(W_b\right)&=1-h_q\left(p_b\right)\;. \label{eq:BSCb}
\end{align}
Using \eqref{eq:v_star}, we see that $\bv_{a\boxast b}\left(y_1,y_2\right)$ are circular shifts of $\tilde\bv_{a\boxast b}\triangleq \left[1-p_t,p_t/(q-1),p_t/(q-1),\dots,p_t/(q-1)\right]^T$, where
\begin{equation}
p_t=p_a+p_b-qp_ap_b/(q-1) \label{eq:p_t}
\end{equation}
so $W_{a\boxast b}$ is a QSC channel with error probability $p_t$, and $I\left(W_{a\boxast b}\right)=1-h_q\left(p_t\right)$. Combined with \eqref{eq:BSCa},\eqref{eq:BSCb} and \eqref{eq:p_t}, this means that for the QSC case, \eqref{eq:Iconv} becomes an equality if $ g (\cdot,\cdot)$ is defined as in \eqref{eq:hat_g_QSC}.
\subsection{Proof of Lemma \ref{lemma:QSCLag3}}
Assume $\bv_a\left(y_1\right)=\left[v_{a,0},v_{a,1},\dots,v_{a,q-1}\right]^T$ and $\bv_b\left(y_2\right)=\left[v_{b,0},v_{b,1},\dots,v_{b,q-1}\right]^T$. Using \eqref{eq:v_star} yields $\bv_{a\boxast b}\left(y_1,y_2\right)=\left[v_{t,0},v_{t,1},\dots,v_{t,q-1}\right]^T$ where
\begin{equation}
v_{t,i}=\sum_{j=0}^{q-1} v_{a,j}v_{b,j-i} \text{ for }i=0,1,\dots,q-1. \label{eq:pti}
\end{equation}
The Lagrangian related to solving the minimization in \eqref{eq:hat_g_def} is
\begin{align}
L&=H\left[\bv_{a\boxast b}\left(y_1,y_2\right)\right]-\lambda_1\left\{H\left[\bv_a\left(y_1\right)\right]-1+G_1\right\}\\
&\qquad -\lambda_2\left\{H\left[\bv_b\left(y_2\right)\right]-1+G_2\right\}-\lambda_3\left(\sum_{i=0}^{q-1}v_{a,i}-1\right)\\
&\qquad-\lambda_4\left(\sum_{i=0}^{q-1}v_{b,i}-1\right)\\
&=
-\sum_{i=0}^{q-1}v_{t,i}\log_q v_{t,i}+\lambda_1\left[1+\sum_{i=0}^{q-1}v_{a,i}\log_q v_{a,i}-G_1\right]\\
&\qquad+\lambda_2\left[1+\sum_{i=0}^{q-1}v_{b,i}\log_q v_{b,i}-G_2\right]\\
&\qquad-\lambda_3\left[\sum_{i=0}^{q-1}v_{a,i}-1\right]-\lambda_4\left[\sum_{i=0}^{q-1}v_{b,i}-1\right] \label{eq:Lagrangian_gen}
\end{align}
and we want to achieve $\left.\partial L\right/ \partial v_{a,i}=\left.\partial L\right/\partial v_{b,i}=0$ for $i=0,1,\dots,q-1$.
By \eqref{eq:pti}, $\left.\partial v_{t,i}\right/\partial v_{a,j}=v_{b,j-i}$ and combining it with \eqref{eq:Lagrangian_gen} and $\sum_{i=0}^{q-1}v_{b,i}=1$ yields
\begin{align}
\frac{\partial L}{\partial v_{a,j}}&=-\frac{1}{\ln q}-\sum_{i=0}^{q-1}v_{b,j-i}\log_q v_{t,i}+\lambda_1\left(\log_q v_{a,j}+\frac{1}{\ln q}\right)\\&\quad-\lambda_3=0\qquad
\forall j\in\left\{0,1,\dots,q-1\right\}\;. \label{eq:Lagrangian_gen_da}
\end{align}
If $W_a$ and $W_b$ are QSC channels, $v_{a,i}=p_a/(q-1)$ and $v_{b,i}=p_b/(q-1)$ for $i\ne 0$, $v_{a,0}=1-p_a$ and $v_{b,0}=1-p_b$. By \eqref{eq:pti}, $v_{t,i}=p_t/(q-1)$ for $i\ne 0$ and $v_{t,0}=1-p_t$, where $p_t$ is defined in \eqref{eq:p_t}. For $j\ne 0$, \eqref{eq:Lagrangian_gen_da} yields
\begin{multline}
-\frac{1}{\ln q} -\frac{p_b}{q-1}\log_q\left[1-p_t\right]-\left(1-\frac{p_b}{q-1}\right)\log_q\frac{p_t}{q-1}\\
+\lambda_1\left(\log_q\frac{p_a}{q-1}+\frac{1}{\ln q}\right)-\lambda_3=0
\end{multline}
and for $j=0$, \eqref{eq:Lagrangian_gen_da} yields
\begin{multline}
-\frac{1}{\ln q} -\left(1-p_b\right)\log_q\left(1-p_t\right)-p_b\log_q\frac{p_t}{q-1}
\\+\lambda_1\left[\log_q\left(1-p_a\right)+\frac{1}{\ln q}\right]-\lambda_3=0\;.
\end{multline}
Now, if $p_a\ne \frac{q-1}{q}$, i.e. $G_1>0$, we have two independent equations, so we have a single possible value for $\lambda_1$ and $\lambda_3$. Combining these equations yields
\begin{align}
\lambda_1=&\left(1-\frac{qp_b}{q-1}\right)\\
&\cdot \log_q\frac{p_t}{(q-1)\left(1-p_t\right)}\left/\log_q\frac{p_a}{(q-1)\left(1-p_a\right)}\right.\\
\lambda_3=&\frac{\frac{p_b}{q-1}\log_q\left(1-p_t\right)+\left(1-\frac{p_b}{q-1}\right)\log_q\frac{p_t}{q-1}}{\log_q\frac{p_a}{q-1}-\log_q\left(1-p_a\right)}\\
&\cdot\log_q\left(1-p_a\right)\\
&-\frac{\log_q\frac{p_a}{q-1}\left[\left(1-p_b\right)\log_q\left(1-p_t\right)+p_b\log_q\frac{p_t}{q-1}\right]
}{\log_q\frac{p_a}{q-1}-\log_q\left(1-p_a\right)}\\
&+\frac{\lambda_1-1}{\ln q}\;.
\end{align}
Similarly, by \eqref{eq:pti}, $\left.\partial v_{t,i}\right/\partial v_{b,j}=v_{a,j+i}$ and combining it with \eqref{eq:Lagrangian_gen} and $\sum_{i=0}^{q-1}v_{a,i}=1$ yields
\begin{multline}
\frac{\partial L}{\partial v_{b,j}}=-\frac{1}{\ln q}-\sum_{i=0}^{q-1}v_{a,j+i}\log_q v_{t,i}+\lambda_2\left(\log_q v_{b,j}+\frac{1}{q}\right)\\-\lambda_4=0\quad\forall j\in\left\{0,1,\dots,q-1\right\}\;. \label{eq:Lagrangian_gen_db}
\end{multline}
If $W_a$ and $W_b$ are QSC channels for $j\ne 0$, \eqref{eq:Lagrangian_gen_db} yields
\begin{multline}
-\frac{1}{\ln q} -\frac{p_a}{q-1}\log_q\left(1-p_t\right)-\left(1-\frac{p_a}{q-1}\right) \log_q \frac{p_t}{q-1}\\+\lambda_2\left(\log_q\frac{p_b}{q-1}+\frac{1}{\ln q}\right)-\lambda_4=0
\end{multline}
and for $j=0$, \eqref{eq:Lagrangian_gen_db} yields
\begin{multline}
-\frac{1}{\ln q} -\left[1-p_a\right]\log_q\left[1-p_t\right]-p_a\log_q\frac{p_t}{q-1}\\
+\lambda_2\left[\log_q\left(1-p_b\right)+\frac{1}{\ln q}\right]-\lambda_4=0\;.
\end{multline}
Now,  if $p_b\ne \frac{q-1}{q}$, i.e. $G_2> 0$, we have  two independent equations, so we have a single possible value for $\lambda_2$ and $\lambda_4$. Combining these equations yields
\begin{align}
\lambda_2=&\left(1-\frac{qp_a}{q-1}\right)\\
&\cdot\log_q\frac{p_t}{(q-1)\left(1-p_t\right)}\left/\log_q\frac{p_b}{(q-1)\left(1-p_b\right)}\right.\\
\lambda_4=&\frac{\frac{p_a}{q-1}\log_q\left(1-p_t\right)+\left(1-\frac{p_a}{q-1}\right)\log_q\frac{p_t}{q-1}}{\log_q\frac{p_b}{q-1}-\log_q\left(1-p_b\right)}\\
&\cdot \log_q\left(1-p_b\right)\\
&-\frac{\log_q\frac{p_b}{q-1}\left\{\left[1-p_a\right]\log_q\left[1-p_t\right]+p_a\log_q\frac{p_t}{q-1}\right\}
}{\log_q\frac{p_b}{q-1}-\log_q\left[1-p_b\right]}\\
&+\frac{\lambda_2-1}{\ln q}\;.
\end{align}
Since we have found $\lambda_1,\dots,\lambda_4$ that solve \eqref{eq:Lagrangian_gen_da} and \eqref{eq:Lagrangian_gen_db} for the case of $W_a$ and $W_b$ being QSC channels, we proved that the QSC case yields a critical point in the Lagrangian related to \eqref{eq:hat_g_def} for any value of $q$.
\comment{
\subsection{Proof of Lemma \ref{lemma:joint}}
By definition, $f\left(\bu_0\right)\triangleq \min_{H\left(\bv\right)\ge 1-G} H\left(\bu_0\star\bv\right)$ and $f\left(\bu_1\right)\triangleq \min_{H\left(\bv\right)\ge 1-G} H\left(\bu_1\star\bv\right)$. Then
\begin{align}
&f\left(\alpha\bu_0+(1-\alpha)\bu_1\right)\\
&\quad=\min_{H\left(\bv\right)\ge 1-G} H\left(\alpha\bu_0\star\bv+(1-\alpha)\bu_1\star\bv\right)\\
&\quad\ge \min_{H\left(\bv\right)\ge 1-G}\left[\alpha H\left(\bu_0\star\bv\right)+(1-\alpha)H\left(\bu_1\star\bv\right)\right]\\
&\quad\ge\alpha \min_{H\left(\bv\right)\ge 1-G} H\left(\bu_0\star\bv\right)+ (1-\alpha)\min_{H\left(\bv\right)\ge 1-G} H\left(\bu_1\star\bv\right)\\
&\quad=\alpha f\left(\bu_0\right)+(1-\alpha)f\left(\bu_1\right)
\end{align}
where the first inequality follows from concavity of $H$, and the added degree of freedom to the minimization yields the second inequality.}
\subsection{Properties of $g_{QSC}$ used in the proof of Lemma \ref{lemma:prop}}
\begin{figure}
		\centering
		\includegraphics[width=\columnwidth]{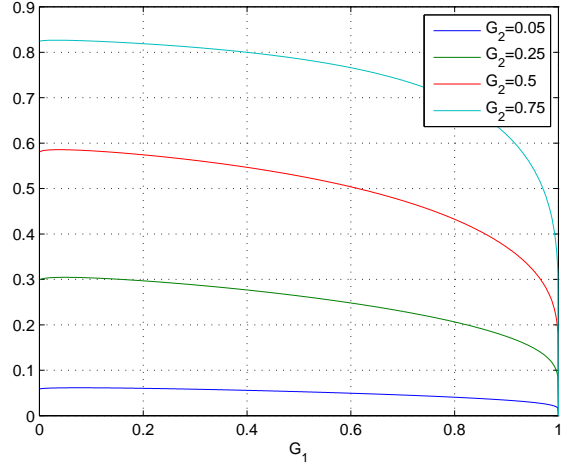}
	\caption{$\frac{\partial g _{QSC}\left(G_1,G_2\right)}{\partial G_1}$ for $q=3$}
	\label{fig:dg}
\end{figure}

By \eqref{eq:hat_g_QSC},
\begin{align}
g_{QSC}\left(1,G_2\right)=1-h_q&\bigg[h_q^{-1}\left(0\right)+h_q^{-1}\left(1-G_2\right)\\ &\quad\left.-\frac{q}{q-1}h_q^{-1} \left(0\right)h_q^{-1}\left(1-G_2\right)\right]\\
=1-h_q&\left[h_q^{-1}\left(1-G_2\right)\right]=G_2
\end{align}
Straightforward calculations show that
\begin{equation}
\frac{\partial g _{QSC}\left(G_1,G_2\right)}{\partial G_1}=\frac{\log_q\left[(q-1)\left(\frac{1}{z}-1\right)\right]\left[1-\frac{q}{q-1}v\right]}{\log_q\left[(q-1)\left(\frac{1}{y}-1\right)\right]} \label{eq:dg}
\end{equation}
where $y=h_q^{-1}\left(1-G_1\right)$, $v=h_q^{-1}\left(1-G_2\right)$ and
$z=y\left(1-v\right)+v\left(1-\frac{y}{q-1}\right)$. These functions are plotted in Figure \ref{fig:dg}. By \eqref{eq:dg}, $\lim_{x\rightarrow 1}\frac{\partial g _{QSC}\left(x,G_2\right)}{\partial x}=0$ (since in this case $y=0$ and $z=v$).
\rem{
\subsection{A proof that $\gamma_l\ge 0$ from Lemma \ref{lemma:largeG1G2}}
In the derivation of Lemma \ref{lemma:largeG1G2} we use \begin{align}
\gamma_l&=\sum_{i=0}^{q-1}\left(v_{a,i}-v_{a,i+l}\right)\log_qv_{a,i}\\
&=\sum_{i=0}^{q-1}v_{a,i}\log_qv_{a,i}-\sum_{i=0}^{q-1}v_{a,i+l}\log_qv_{a,i}\\
&=\sum_{i=0}^{q-1}v_{a,i+l}\log_qv_{a,i+l}-\sum_{i=0}^{q-1}v_{a,i+l}\log_qv_{a,i}\\
&=\sum_{i=0}^{q-1}v_{a,i+l}\log_q\left(v_{a,i+l}/v_{a,i}\right)
\end{align}
where the third equality follows from the circular nature of the first sum.
This result can be seen as the Kullback-Lieber divergence between a circular shift of $\bv_a\left(y_1\right)$ and $\bv_a\left(y_1\right)$ itself. Since Kullback-Lieber divergence is non-negative, $\gamma_l\ge 0$.
}

\subsection{A proof that $(x+y)\ln(x+y)\approx x\ln x+y\ln y$ for small positive $x,y$}
We are going to prove that $$1\le\frac{x\ln x+y\ln y}{(x+y)\ln(x+y)}\le1-\frac{\ln2}{\ln(x+y)}$$ so $$\lim_{x,y\rightarrow 0}\frac{x\ln x+y\ln y}{(x+y)\ln(x+y)}=1\;.$$ 
First, since $-x\ln x$ is concave, 
\begin{multline}
\frac{-x\ln x-y\ln y}{2}\le -\left(\frac{x+y}{2}\right)\ln\left(\frac{x+y}{2}\right)\\
=-\left(\frac{x+y}{2}\right)\ln(x+y)+\left(\frac{x+y}{2}\right)\ln2
\end{multline}
so, dividing both sides by $-0.5(x+y)\ln(x+y)$ yields $$\frac{x\ln x+y\ln y}{(x+y)\ln(x+y)}\le1-\frac{\ln2}{\ln(x+y)}\;.$$
For the other direction we must prove that $$-x\ln x-y\ln y\ge -(x+y)\ln(x+y)\;.$$ It is equivalent to
$$x\left[\ln(x+y)-\ln x\right]\ge y\left[\ln y-\ln(x+y)\right]$$
Since $\ln$ is an increasing function, the left hand side of the inequality above is positive, and the right hand side is negative, so it is a true statement.

Note that for $q$ variables (instead of $2$) the first half of the proof is similar, using $q$ instead of $2$, and the second half is modified using $q-1$ induction steps, one for each sum.
\subsection{Proof of Lemma \ref{lemma:one-sol}}
Define $f(x)\triangleq x\cdot\frac{\partial g \left(x,G_2\right)}{\partial x}- g \left(x,G_2\right)$. We wish to prove that $f(x)=0$ has exactly one solution that satisfies $\frac{\partial^2 g \left(x,G_2\right)}{\partial x^2}<0$.
First, $f'(x)=x\cdot\frac{\partial^2 g \left(x,G_2\right)}{\partial x^2}$. Since there exists $x^*$ s.t. $\frac{\partial^2 g \left(x,G_2\right)}{\partial x^2}$ is positive for $x<x^*$ and negative for $x>x^*$ (See Property \ref{prop:g(G1,G2)}), $f(x)$ is increasing for $x<x^*$ and decreasing for $x>x^*$. Combining this with $f(0)=0$ yields that $f(x)>0$ for $0<x\le x^*$. Lemma \ref{lemma:prop} shows that $\lim_{x\rightarrow1}\frac{\partial g \left(x,G_2\right)}{\partial x}=0$ and $ g \left(1,G_2\right)=G_2$, so $\lim_{x\rightarrow 1}f(x)=-G_2$. Since $f\left(x^*\right)>0$, $\lim_{x\rightarrow 1}f(x)<0$, and $f(x)$ is decreasing for $x^*\le x\le 1$, $f(x)=0$ has exactly one solution for $x^*<x\le 1$. The only other solution to $f(x)=0$ is $x=0$, and in this point $\frac{\partial^2 g \left(x,G_2\right)}{\partial x^2}>0$.
\rem{
\subsection{A plot of $\frac{1}{k}\log_2 L_k$}
The values of  $\frac{1}{k}\log_2 L_k$ as a function of $k$ for $q=3$ and $f_0(z)=\left(0.26x^2+1\right)x^{0.8}(1-x)^{0.6}$ 
are plotted in Figure \ref{fig:Lkx3}. Similarly to \cite[Fig. 3]{goldin2013improved}, the plot suggests it is a decreasing function, but it decreases slowly for $k>100$, so using $\rho=-\frac{1}{100}\log_2 L_{100}$ is a good choice for deriving Proposition \ref{theo:hassani17}.
\begin{figure}
\centering
\includegraphics[width=\columnwidth]{Lkx31.eps}
\caption{A plot of $\frac{1}{k}\log_2 L_k$ as a function of $k$ for $q=3$ and $f_0(x)=\left(0.26x^2+1\right)x^{0.8}(1-x)^{0.6}$} \label{fig:Lkx3}
\end{figure}
}
\end{document}

%% file: shortcuts.tex
\newcommand{\rem}[1]{}

\newtheorem{lemma}{Lemma}

\newtheorem{proposition}{Proposition}
\newtheorem{property}{Property}

\newcommand{\bre}{\begin{equation}}
\newcommand{\ere}{\end{equation}}

\newcommand{\ee}\]
\newcommand{\bra}{\begin{eqnarray}}
\newcommand{\era}{\end{eqnarray}}
\newcommand{\bfg}{\begin{figure}[hbtp]}
\newcommand{\efg}{\end{figure}}
\newcommand{\bver}{\begin{verbatim}}
\newcommand{\ever}{\end{verbatim}}
\newcommand{\bit}{\begin{itemize}}
\newcommand{\eit}{\end{itemize}}
\newcommand{\ben}{\begin{enumerate}}
\newcommand{\een}{\end{enumerate}}

\newcommand{\bepsilon}{\boldsymbol\epsilon}

\newcommand{\given}{\: | \:}

\newcommand{\baa}{\begin{eqnarray*}}
\newcommand{\eaa}{\end{eqnarray*}}

\newcommand{\bu}{{\bf u}}
\newcommand{\bv}{{\bf v}}

\newcommand{\hW}{\hat{W}}

\newcommand{\rE}{{\rm E}}

\newcommand{\defined}{\triangleq}

\def\argmax{\mathop{\rm argmax}}

\def\defined{\: {\stackrel{\scriptscriptstyle \Delta}{=}} \: }

\newfont{\boldlarge}{msbm10 scaled 1100}

\newcommand{\comment}[1]{}

\renewcommand{\thesection}{\Roman{section}}

\newlength{\tmpbigbar}
